\documentclass[a4paper,11pt]{article}
\usepackage{pos}
\usepackage{braket}

\title{Complex poles of Landau-gauge QCD propagators and general properties}

\author*[a]{Yui Hayashi}
\author[a,b]{Kei-Ichi Kondo}

\affiliation[a]{
Department of Physics, Graduate School of Science and Engineering, Chiba University, Chiba 263-8522, Japan
}

\affiliation[b]{
Department of Physics, Graduate School of Science, Chiba University, Chiba 263-8522, Japan
}

\emailAdd{yhayashi@chiba-u.jp}
\emailAdd{kondok@faculty.chiba-u.jp}

\abstract{We study analytic structures of the gluon, quark, and ghost propagators in the Landau-gauge QCD and general properties from the existence of unusual singularities. First, we investigate analytic structures of the QCD propagators using the massive Yang-Mills model, in which the one-loop gluon and ghost propagators are in good agreement with the numerical lattice results in the Landau gauge. We find that both gluon and quark propagators in this model have complex poles that invalidate the usual spectral representation. Second, we discuss general properties of propagators in the presence of such complex singularities, especially on the positivity and locality. Finally, we consider a possible quantum mechanical interpretation and implications on a confinement mechanism.}

\FullConference{%
 The 38th International Symposium on Lattice Field Theory, LATTICE2021
  26th-30th July, 2021
  Zoom/Gather@Massachusetts Institute of Technology
}

\begin{document}
\maketitle

\section{Introduction}

The analytic structure of a propagator represents relevant states and spectrum.
For example, for a physical case, the K\"all\'en--Lehmann spectral representation indicates that singularities on the complex momentum plane correspond to states non-orthogonal to the state $\phi(0)  \ket{0}$:
\begin{align}
D(k^2) &= \int_0 ^\infty d \sigma^2 \frac{\rho(\sigma^2)}{\sigma^2 - k^2}, \label{eq:KL}\\
   \theta(k_0) \rho(k^2) 
&:=  (2\pi)^{d} \sum_{n }  |\langle 0 | \phi(0) | P_n \rangle|^2 \delta^D(P_n-k).
\end{align}
Therefore, we can probe states and spectra related to the field by looking into the propagator.
For understanding fundamental aspects of QCD, for example, confinement, investigating analytic structures of the QCD propagators for confined particles would be useful.

For many years, the gluon, ghost, and quark propagators in the Landau gauge have been extensively studied by both lattice and continuum methods.
Based on this progress, more recent attention has focused on analytic structures of the QCD propagators.

Remarkably, recent analyses of independent approaches, e.g., numerical reconstruction techniques from Euclidean data \cite{BT2019, Falcao:2020vyr}, models of massivelike gluons \cite{Siringo16, HK2018, Hayashi:2020few}, and the ray technique of the Dyson-Schwinger equation \cite{SFK12,Fischer-Huber}, suggest that the Landau-gauge gluon propagator has \textit{complex singularities}, namely unusual singularities invalidating the K\"all\'en-Lehmann spectral representation.

However, the interpretation of complex singularities has been less studied. There are only old works \cite{Stingl} discussing this subject heuristically.
Hence, we study general properties of propagators with such singularities \cite{Hayashi:2021nnj, Hayashi:2021jju}.

In this presentation, after reviewing how we explore the analytic structures, we show general properties and interpretation of complex singularities.

\section{Analytic structures of the propagators}

Let us briefly describe how the analytic structures are investigated and review the model of massivelike gluons as an example.

\subsection{Methodology}

\begin{figure}[t]
    \centering
    \includegraphics[width =  0.7 \linewidth]{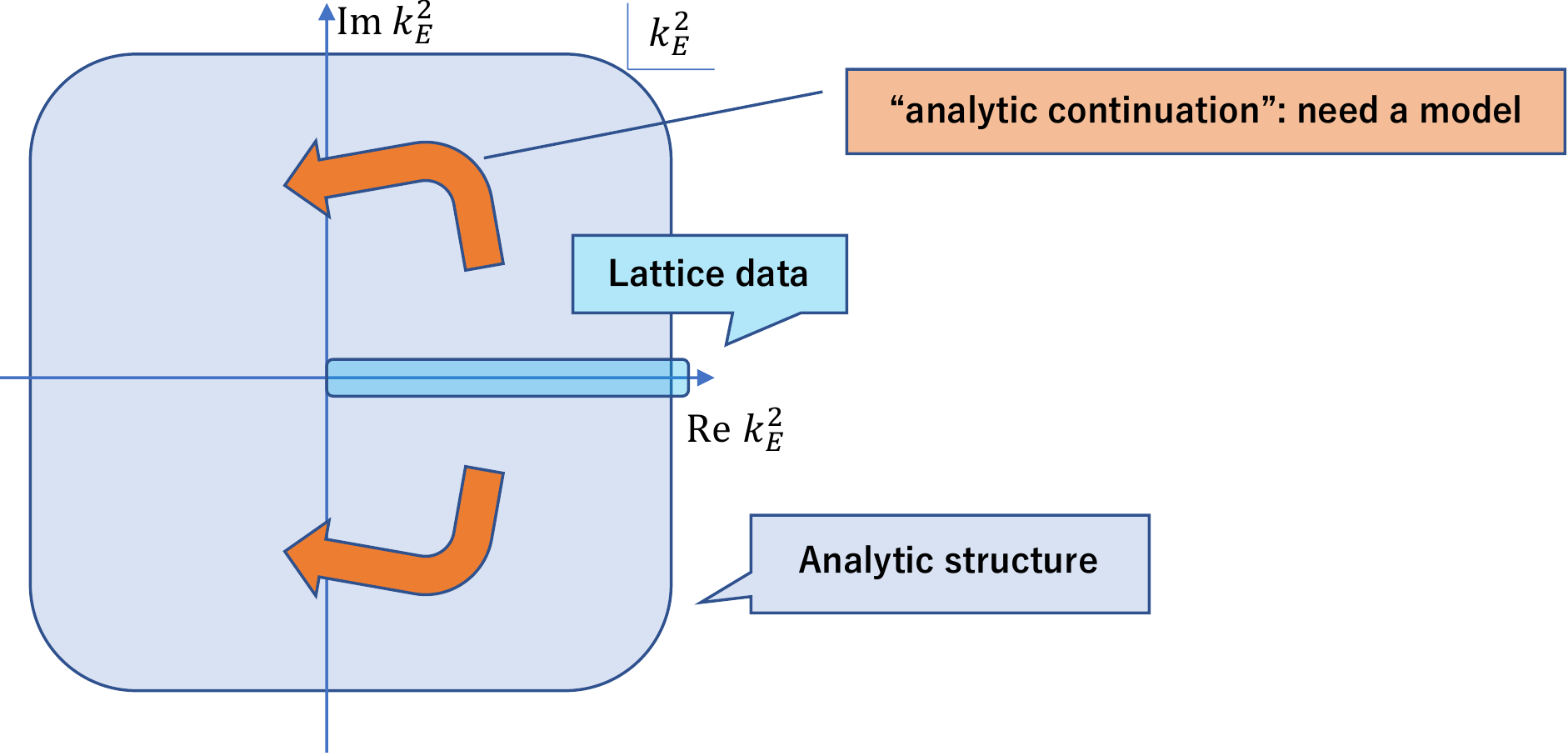}
    \caption{Conceptual picture describing how the analytic structures are investigated in many works. Note that this analytic structure is defined on the complexified Euclidean momentum plane.}
    \label{fig:concept_analytic}
\end{figure}

Our aim in this section is to investigate analytic structures of the propagators using the Euclidean lattice data through some analytic continuation.
Note that such analytic continuation from finite lattice data is an ill-posed problem; the uniqueness of an analytic continuation is not guaranteed.
Nevertheless, analytic structures can be speculatively studied by using a model satisfying:
\begin{itemize}
    \item It has some theoretical backgrounds or motivations.
    \item It reproduces the propagators in good agreement with lattice data.
    \item Its (analytically-continued) propagators on the complex $k^2$-plane can be computed.
\end{itemize}
If we have such a model, the model propagators provide possible analytic structures of the propagators.
See Fig.~\ref{fig:concept_analytic} for this procedure.

We emphasize that what we examine here is the analytic structure on the complexified \textit{Euclidean} momentum plane.
Although trivial, this is an important point for rigorous consideration which is discussed in the next section.

\subsection{Example: analytic structures by the massive Yang-Mills model}

As mentioned in the Introduction, there are many studies based on the above methodology.
As an example of these studies, let us review the method by the massive Yang-Mills model, which is the Landau-gauge limit of the Curci-Ferrari model \cite{HK2018, Kondo:2019rpa}.
See \cite{Pelaez:2021tpq} for a review of this approach to the infrared Yang-Mills theory.

This model consists of the usual Landau-gauge Faddeev-Popov Lagrangian and a naive gluon mass term:

\begin{align}
{\cal L}_{mYM} &=  \frac{1}{4} F^A_{\mu \nu} F^{A}_{\mu \nu} + i B^A \partial_\mu A^A_\mu +  \bar{c}^A \partial_\mu {\cal D}_\mu[A]^{AB} c^B + \frac{1}{2} M^2 A_\mu ^A A_\mu ^A .
\end{align}

Let us mention motivations for adding the mass term:
(1) Minimal deformation to mimic the massivelike behavior.
(2) Gribov ambiguity. 
In the lattice calculations, the Landau gauge is realized by trying to minimize the functional $\int d^4x~ A_\mu ^A A_\mu ^A$.
Therefore, there should be some effect suppressing $\int d^4x~ A_\mu ^A A_\mu ^A$ compared to the standard Faddeev-Popov Lagrangian.
(3) Dimension-two gluon condensate.
There is an argument that the dimension-two gluon operator $A_\mu ^A A_\mu ^A$ condensates.
After this condensation, the quadratic term will appear.
Therefore, the effective mass term could be understood as a most tractable and infrared-relevant way to include these effects.

\begin{figure}[t]
    \begin{tabular}{cc}
      \begin{minipage}[t]{0.4 \hsize}
        \centering
   \includegraphics[width= \linewidth ]{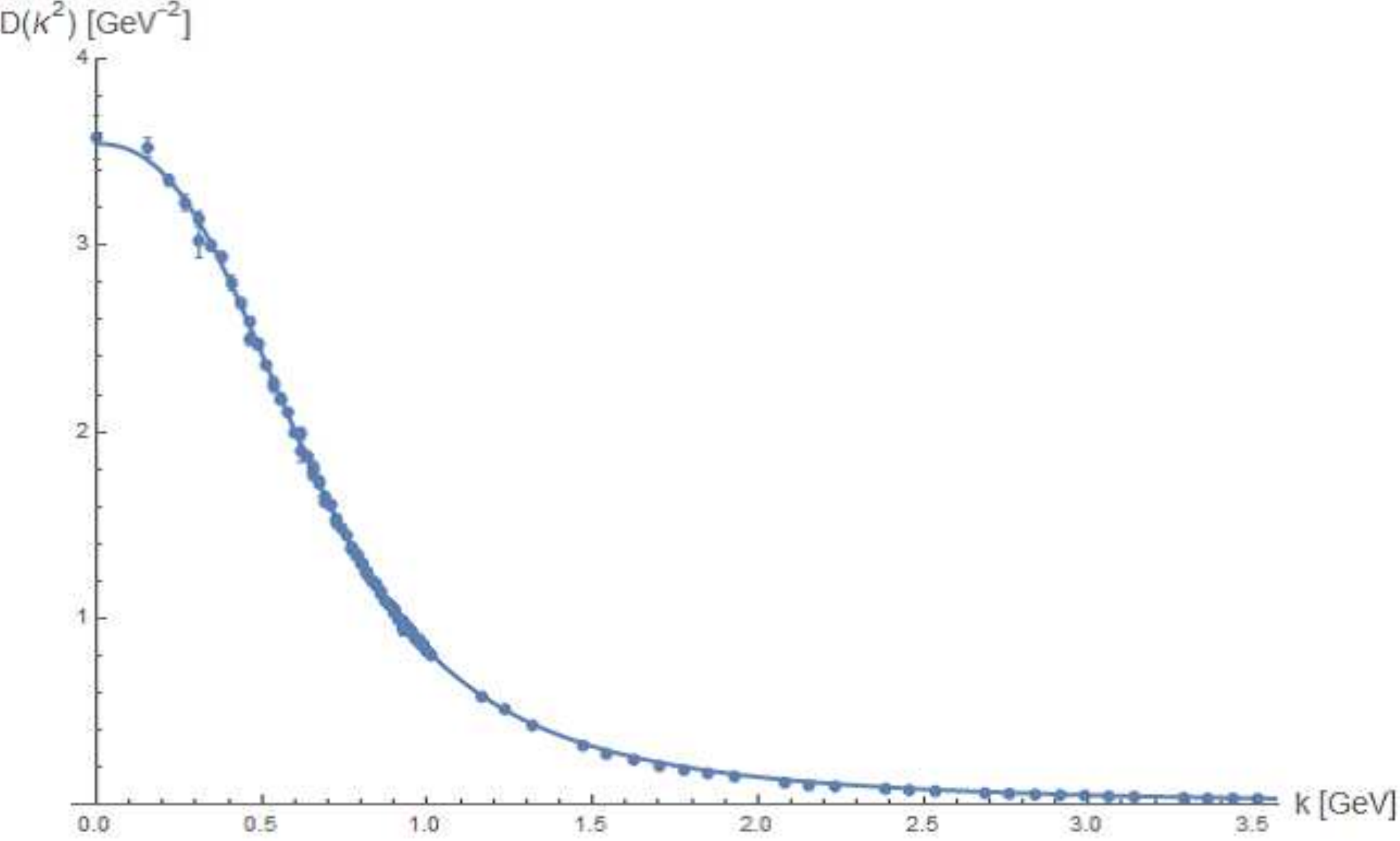}
      \end{minipage} &
      \begin{minipage}[t]{0.59\hsize}
        \centering
   \includegraphics[width=  \linewidth]{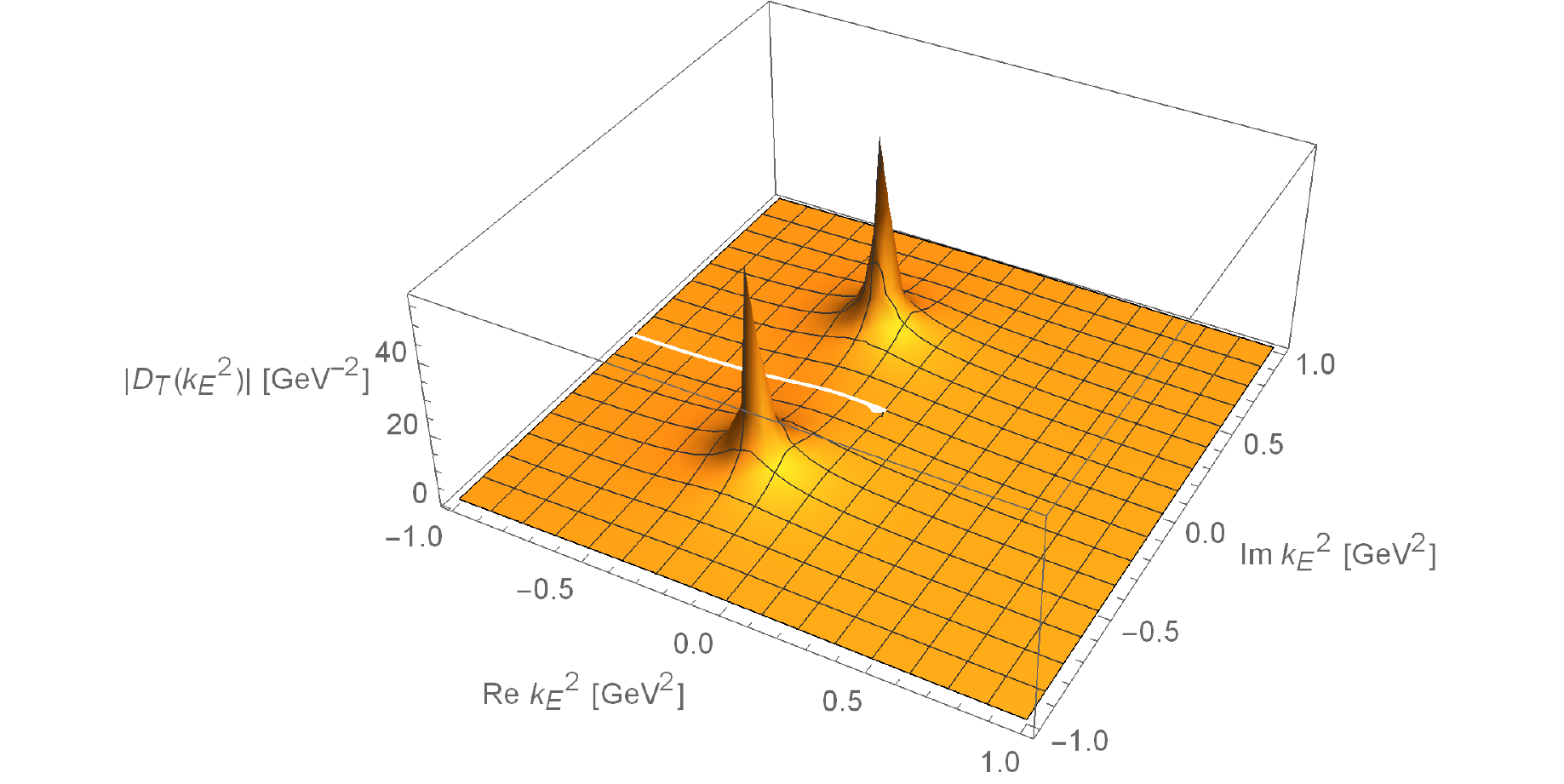}
      \end{minipage}
    \end{tabular}
    \caption{(Left panel) the one-loop gluon propagator at the best-fit parameters and lattice data \cite{Duarte:2016iko} of the gluon propagator in the $SU(3)$ Yang-Mills theory.   (Right panel) Modulus of the gluon propagator on the complex $k_E^2$-plane \cite{HK2018}. Complex poles are found at $k_E^2 = -0.23 \pm 0.42i \mathrm{~GeV}^2$.}
    \label{fig:fitted_propagator}
  \end{figure}

As an advantage of this model, by choosing the parameters\footnote{With the ``infrared-safe'' renormalization condition \cite{Tissier:2011ey},
\begin{align*}
 \begin{cases}
 Z_A Z_C Z_{M^2} = 1,~~  Z_g \sqrt{Z_A} Z_C  = 1 \\
 \Gamma_{A}^{(2)} (k_E = \mu) = \mu^2 + M^2, ~~ \Gamma_{gh}^{(2)}(k_E = \mu) = \mu^2,
 \end{cases}
\end{align*}
at the renormalization scale $\mu = 1 \mathrm{~GeV}$, the best-fit parameters are $g = 4.1, ~~ M= 0.45 \mathrm{~GeV}$. See Sec.~III~E of \cite{Kondo:2019rpa} for details.
As another advantage of the model, in this renormalization condition, the gauge running coupling is always `small' for all scales, which partially justifies the use of the perturbation theory \cite{Tissier:2011ey}.}, the gluon and ghost propagators of this model exhibit striking agreement with lattice results in the one-loop level.
See the left panel of Fig.~\ref{fig:fitted_propagator} for the gluon propagator.

We can easily analytically continue the one-loop expression of the gluon propagator and find complex poles (right panel of Fig.~\ref{fig:fitted_propagator})\footnote{Incidentally, with two dynamical quarks, we have a similar conclusion for the gluon and quark propagators \cite{Hayashi:2020few}, while the ghost propagator has no complex poles.}.
In summary, the massive Yang-Mills model has some theoretical backgrounds and reproduces the lattice propagators.
The modeled gluon propagator has unusual singularities, complex poles.

Obviously, the complex poles invalidate the K\"all\'en--Lehmann spectral representation (\ref{eq:KL}).
Such singularities are beyond the standard formalism of QFT. Therefore, for an interpretation of this result, we have to consider complex singularities carefully.

\section{General properties of complex singularities}

Let us move on to the main topic.
Here, we show some rigorous results for propagators with complex singularities and how complex singularities are realized in state spaces.

The purpose here is to obtain an interpretation of complex singularities.
As emphasized in the previous section, complex singularity is a property of the Euclidean propagator: we need a ``reconstruction'' to this end.

\begin{figure}[t]
    \centering
    \includegraphics[width =  0.7 \linewidth]{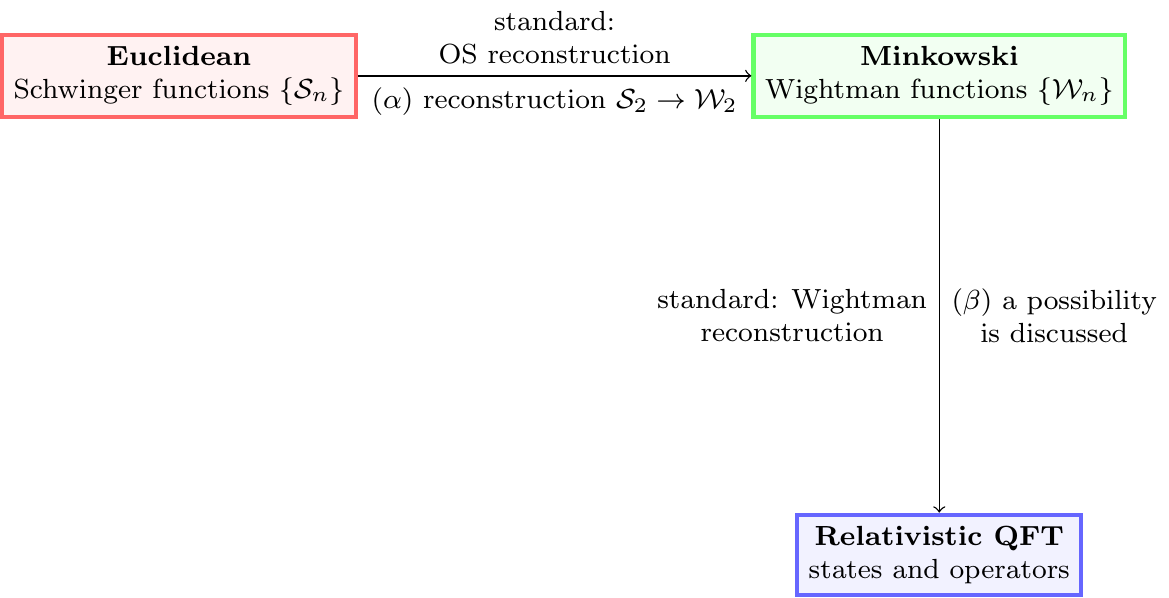}
    \caption{Standard reconstruction procedure and contents of our study ($\alpha$) and ($\beta$). Taken from \cite{Hayashi:2021nnj}.}
    \label{fig:reconstruction}
\end{figure}

To clarify the starting point, let us briefly summarize how we reconstruct quantum theories from Euclidean field theories in the standard case \cite{OS73, OS75} (Fig.~\ref{fig:reconstruction}).
We start with a set of Euclidean correlation functions, called Schwinger functions.
If these Schwinger functions satisfy the Osterwalder-Schrader axioms, e.g., Euclidean invariance, reflection positivity, and so on, we can reconstruct the Wightman functions on the Minkowski spacetime by the analytic continuation in the position space.
Subsequently, by the Wightman reconstruction, we can obtain a quantum theory written in terms of states and operators from the Wightman functions.
Note that this reconstruction is implicitly performed in many cases, since almost all nonperturbative methods are developed in Euclidean field theories.

In this section, when complex singularities are present, we look into this reconstruction procedure:

\begin{itemize}
    \item [($\alpha$)] Making an analytic continuation from the Schwinger function [$S(\vec{\xi} , \xi_4) = W(- i \xi_4, \vec{\xi})$] to the Wightman function $W(\xi^0, \vec{\xi})$ on the Minkowski spacetime.
    \item [($\beta$)]  Finding a state-space structure reproducing the reconstructed Wightman function $W(\xi^0, \vec{\xi})$.
\end{itemize}

We make some technical assumptions on the analytic structure: (1) boundedness of complex singularities in $|k_E^2|$ in the complex $k_E^2$ plane, (2) holomorphy of $D(k_E^2)$ in a neighborhood of the real axis except for the timelike ($k_E^2 < 0$) singularities, (3) some regularity of discontinuity on the timelike axis.

\textbf{Results ($\alpha$).} For the procedure ($\alpha$), we obtain the following rigorous results \cite{Hayashi:2021jju, Hayashi:2021nnj}.
\begin{enumerate}
\renewcommand{\labelenumi}{(\Alph{enumi})}
    \item The reflection positivity
    is violated for the Schwinger function.
    \item The holomorphy of the Wightman function $W(\xi - i \eta)$ in the tube $\mathbb{R}^4 - i V_+$ and the existence of the boundary value as a distribution are still valid, where $V_+$ denotes the (open) forward light cone. Thus, we can reconstruct the Wightman function from the Schwinger function.
    \item The temperedness and the positivity condition are violated for the reconstructed Wightman function. The spectral condition is never satisfied since it requires the temperedness as a prerequisite.
    \item The Lorentz symmetry and spacelike commutativity are kept intact.
\end{enumerate}

For detailed proofs of these assertions, see \cite{Hayashi:2021nnj}. Here, we only show intuitive derivations with a simple example.

\textbf{Example.}
Let us see these properties with a simple example: one pair of complex conjugate poles (e.g., the typical Gribov-Zwanziger fit),
\begin{align}
    D(k_E^2) = \frac{Z}{k_E^2 + M^2} + \frac{Z^*}{k_E^2 + (M^*)^2},
\end{align}

(B) The Schwinger function reads
\begin{align}
    S (\vec{\xi}, \xi_4) = \int \frac{d^3 \vec{k}}{(2 \pi)^3} e^{i\vec{k} \cdot \vec{\xi}} \left[ \frac{Z}{2 E_{\vec{k}} } e^{- E_{\vec{k}} |\xi_4|} +  \frac{Z^*}{2 E_{\vec{k}}^*} e^{- E_{\vec{k}}^* |\xi_4|}
    \right], \label{eq:simple_complex_poles_Schwinger}
\end{align}
where $E_{\vec{k}} = \sqrt{\vec{k}^2 +M^2}$ is a branch of $\operatorname{Re} E_{\vec{k}} > 0$.
We consider the analytic continuation of the Wightman function starting from the imaginary-time data $S(\vec{\xi} , \xi_4) = W(- i \xi_4, \vec{\xi})$.
The analytic continuation to the tube $\mathbb{R}^4 - iV_+$ can be simply given by the integral representation.
\begin{align}
    W&(\xi - i\eta) = \int \frac{d^3 \vec{k}}{(2 \pi)^3} e^{i\vec{k} \cdot (\vec{\xi} - i \vec{\eta})} \left[ \frac{Z}{2 E_{\vec{k}} } e^{- i E_{\vec{k}} (\xi^0 - i \eta^0)} +  \frac{Z^*}{2 E_{\vec{k}}^*} e^{- i E_{\vec{k}}^* (\xi^0 - i \eta^0)} 
    \right]. \label{eq:simple_complex_poles_hol_Wightman}
\end{align}
Since the integrand decreases rapidly in $|\vec{k}|$ for $\eta \in V_+$, this expression is holomorphic in the tube $\xi - i \eta \in \mathbb{R}^4 - iV_+$. 
We can take the ``limit'' $\eta \rightarrow 0~ (\eta \in V_+)$ of (\ref{eq:simple_complex_poles_hol_Wightman}) as a distribution. A subtle point of this limit is the integral over $\vec{k}$, which is just the Fourier transformation and can be defined properly as a distribution.
We thus have the reconstructed Wightman function on the Minkowski spacetime:
\begin{align}
    W&(\xi) = \int \frac{d^3 \vec{k}}{(2 \pi)^3} e^{i\vec{k} \cdot \vec{\xi} } \left[ \frac{Z}{2 E_{\vec{k}} } e^{- i E_{\vec{k}} \xi^0} +  \frac{Z^*}{2 E_{\vec{k}}^*} e^{- i E_{\vec{k}}^* \xi^0} 
    \right]. \label{eq:Wightman}
\end{align}

(C) Since $E_{\vec{k}}$ is complex, $W(\xi)$ exhibits an exponential growth in $\xi^0$ beyond polynomial growth. Therefore, the Wightman function on the Minkowski spacetime violates the temperedness.
We can also show
\begin{align}
    \mathrm{(Positivity) \Rightarrow (Temperedness)}.
\end{align}
Intuitively, this can be understood as follows.
\begin{enumerate}
    \item The positivity of $W(\xi)$ corresponds to the positivity of $\{ \phi(x) \ket{0} \}_{x \in \mathbb{R}^4}$.
    \item The assumed translational invariance of the two-point function corresponds to the unitarity of the translation operator defined on this sector: $U(a) \phi(x) \ket{0} := \phi(x + a) \ket{0}$.
\end{enumerate}
Then, we have an $a$-independent ``upperbound'' on $|W(a)| = |\braket{0|\phi(0) U(-a) \phi(0)|0} | \leq \braket{0|\phi(0) \phi(0)|0}$, which will imply that $W(a)$ is tempered\footnote{Of course, since $W(\xi)$ is a distribution, the upperbound does not generally exist. Nevertheless, we can also prove the claim rigorously in the same spirit.}.

(A) 
Similarly, the reflection positivity contradicts the nontemperedness. This can be shown by repeating a part of the Osterwalder-Schrader reconstruction \cite{OS73} from Schwinger functions to Wightman functions.

(D)
We can show the Lorentz invariance from the Euclidean rotational invariance of the imaginary-time data. The holomorphic Wightman function $W(\xi - i\eta)$ is invariant under infinitesimal Euclidean rotations because of the uniqueness of the analytic continuation. It is also invariant under the complexified versions, infinitesimal complex Lorentz transformations. Therefore, by a restriction to the Minkowski spacetime, the Lorentz invariance will remain.
We can also explicitly check the Lorentz invariance of the expression (\ref{eq:simple_complex_poles_hol_Wightman}) by contour deformation.

For the case with a single scalar field, the spacelike commutativity [$W(\xi) = W(-\xi)$ for spacelike $\xi$] is an immediate consequence from the Lorentz invariance.
For general cases, this follows from the permutation symmetry of the Schwinger function and the complex Lorentz covariance of the holomorphic Wightman function.

\medskip
\textbf{Result ($\beta$).}  For the procedure ($\beta$), we find \cite{Hayashi:2021jju, Hayashi:2021nnj}:
\begin{enumerate}
\renewcommand{\labelenumi}{(\Alph{enumi})}
\setcounter{enumi}{4}
    \item Complex singularities can be realized in indefinite-metric QFTs and correspond to pairs of zero-norm eigenstates of complex energies.
\end{enumerate}

This claim can be understood as follows.
States with complex conjugate eigenvalues of a hermitian Hamiltonian can be realized by zero-norm pairs in an indefinite metric state space:
\begin{align*}
    (\ket{E}, \ket{E^*})~
    \begin{cases}
    H \ket{E} = E \ket{E}, ~~~ H \ket{E^*} = E^* \ket{E^*} \\
    \braket{E|E} = \braket{E^*|E^*} = 0,~~ \braket{E|E^*} \neq 0
    \end{cases}
\end{align*}
If such states are present, the Wightman function has exponential factors,
\begin{align*}
    \braket{0|\phi(t) \phi(0)|0} &\supset (\braket{E^*|E})^{-1} e^{- i E t} \braket{0|\phi(0)|E} \braket{E^*|\phi(0)|0} \\
    & ~~~ + (\braket{E|E^*})^{-1} e^{- i E^* t} \braket{0|\phi(0)|E^*} \braket{E|\phi(0)|0}.
\end{align*}
By preparing such states for all momentum $\vec{p}$, we can reproduce the Wightman function reconstructed from complex poles (\ref{eq:Wightman}).

Finally, let us add some comments on these results.

\textbf{Nontemperedness}.
The exponential growth of $W(\xi)$, which stems from complex-energy states, strongly suggests that asymptotic states of the field are ill-defined.
Such states in the full state space should be excluded from the physical state space before taking the asymptotic limit ($\xi^0 \rightarrow \infty$) through some confinement mechanism.
In this sense, the complex singularities might be considered as a signal of confinement.

\textbf{BRST symmtery}.
A promising way to eliminate such states is the Kugo-Ojima quartet mechanism \cite{Kugo:1979gm} by the BRST symmetry.
It can be argued that the existence of complex singularities in a propagator of the gluon-ghost composite operator is a necessary condition for this scenario \cite{Hayashi:2021jju}.
Testing this condition would be interesting.
Incidentally, in a similar point of view, the Bethe-Salpeter equation for the gluon-ghost bound state was discussed in {\cite{Alkofer:2011pe}}.

\textbf{Locality.}
Complex singularities are often associated with non-locality in some literature since they are beyond the usual formalism of local QFTs.
However, to the best of our knowledge, the only axiomatic way to impose locality is the spacelike commutativity.
Consequently, from the result (D), complex singularities themselves do not necessarily lead to non-locality.

\textbf{Wick rotation.}
Since the reconstructed Wightman function (\ref{eq:Wightman}) diverges in both limits $\xi^0 \rightarrow \pm \infty$, the time-ordered propagator cannot be Fourier-transformed either.
Thus, the naive inverse Wick rotation in the momentum space $k_E^2 \rightarrow - k^2$ cannot be applied in the presence of complex singularities\footnote{Note that, if we instead adopted the inverse Wick rotation $k_E^2 \rightarrow - k^2$ ad hoc, the Euclidean field theory could not be interpreted as an imaginary-time formalism, and the hermiticity of the Hamiltonian would be violated.}.
This inverse Wick rotation is not a fundamental assumption but a consequence of the spectral representation, which is invalidated here.

\section{Summary}

We have reviewed how complex singularities are suggested in many works and shown general properties when such singularities exist. 
As a result, the Lorentz symmetry and locality are kept, while the temperedness and the positivity condition are violated.
Moreover, complex singularities can be understood as pairs of zero-norm eigenstates of complex energies in indefinite-metric QFTs, which should be confined.

In conclusion, although complex singularities are beyond the standard formalism of QFT, they can appear in indefinite-metric QFTs (such as gauge theories in Lorentz covariant gauges) and can be understood as confined states.

\acknowledgments
Y.~H. is supported by JSPS Research Fellowship for Young Scientists Grant No.~20J20215, and K.-I.~K. is supported by Grant-in-Aid for Scientific Research, JSPS KAKENHI Grant (C) No.~19K03840.

\end{document}